\documentstyle[epsf,preprint,eqsecnum,aps,rotate]{revtex}

\global\firstfigfalse

\begin{document}

\title{\bf Observation of the GZK Cutoff by the HiRes Experiment}

\author{G.~B.~Thomson$^a$\footnote{{\bf e-mail}: thomson@physics.rutgers.edu}
\\for the High Resolution Fly's Eye (HiRes) Collaboration \\
$^a$ \small{\em Rutgers University} \\
\small{\em Department of Physics and Astronomy\\Piscataway, NJ 08854, USA}
}

\date{September 13, 2006}
\maketitle

\begin{abstract}

The High Resolution Fly's Eye (HiRes) experiment has observed the GZK
cutoff.  HiRes' measurement of the flux of cosmic rays shows a sharp
suppression at an energy of $6 \times 10^{19}$ eV, exactly the
expected cutoff energy.  We observe the ``ankle'' of the cosmic ray
spectrum as well, at an energy of $4 \times 10^{18}$ eV.  However we
cannot claim to observe the third spectral feature of the ultrahigh
energy cosmic ray regime, the ``second knee''.  We describe the
experiment, data collection, analysis, and estimate the systematic
uncertainties.  The results are presented and the calculation of a
five standard deviation observation of the GZK cutoff is described.  

\end{abstract}

\section{Introduction:  The GZK Cutoff}

K. Greisen \cite{g} and G. Zatsepin and V. Kuzmin \cite{zk} predicted
in 1966 that there would be an end to the cosmic ray spectrum.  Caused
by the production of pions in collisions between cosmic ray protons
and photons of the cosmic microwave background radiation (CMBR), this
strong suppression is called the GZK cutoff and should occur at about
$6 \times 10^{19}$ eV.  This energy is set by the mass and width of
the $\Delta$ resonance and the temperature of the CMBR.  Several
experiments \cite{early,pravdin} whose exposures were too small to
measure the flux of cosmic rays at the cutoff energy nevertheless
observed one event each above $10^{20}$ eV, raising questions about
the validity of Greissen's and Zatsepin and Kuzmin's prediction.

The Akeno Giant Air Shower Array (AGASA) was the first experiment
large enough to measure the flux of cosmic rays at $10^{20}$ eV.
AGASA consisted of 111 scintillation counters placed in an array of
spacing about 1 km, and completed its data collection in 2004.  They
found that the spectrum continued unabated beyond the GZK cutoff
energy: their main evidence consisted of 11 events above $10^{20}$ eV
\cite{shinozaki}.  These results formed a serious challenge to the
expectation that the cosmic ray flux should be cut off.  Since the
prediction of the existence of the cutoff is based on accepted physics
principles this was a truly startling result.

The High Resolution Fly's Eye (HiRes) experiment collected data for
nine years (1997-2006), and now has a much higher exposure than AGASA
near the cutoff energy.  HiRes is a fluorescence experiment; i.e., it
observes cosmic ray showers by collecting the fluorescence light
emitted by nitrogen molecules excited by shower particles.  By using
this different technique HiRes has much better resolution and better
control of systematic uncertainties than AGASA.  A fluorescence
detector collects much more information about cosmic ray showers (for
example it sees the shower development), measures the energy more
accurately, and because it depends only on what goes on in the center
of the shower one can simulate a fluorescence detector's performance
very accurately.  This is an important point that has not been lost on
new experiments being constructed: the Auger and Telescope Array
experiments have both fluorescence detectors and ground arrays.

HiRes sees the GZK cutoff.  In this article we describe the
measurement of the flux of cosmic rays in which this observation
occurs, show the cosmic ray spectrum that we observe, and discuss
of our estimates of systematic uncertainties.

\section{The HiRes Experiment}

The HiRes experiment has been described previously \cite{expt}.  We
have two detectors located atop desert mountains in west-central Utah,
separated by 12.6 km, that operate on clear, moonless nights.  The
older detector, called HiRes-I, consists of a ring of spherical
mirrors that collect the fluorescence light.  Each mirror focuses the
light on a 16x16 array of photomultiplier tubes.  As the shower comes
down across the sky the image passes over various photomultiplier
tubes.  A sample-and-hold readout system is used to save the timing
and pulse height information from each tube.  The mirrors have an
active area of 4.2 m$^2$, and the 21 mirrors cover elevation angles
from 3 to 17 degrees and almost the full azimuthal angle.  Each
phototube subtends about 1 square degree of sky.

Our newer detector, called HiRes-II, has 42 mirrors arranged in two
rings that cover from 3 to 31 degrees in elevation and almost the full
azimuthal angle.  The front end of the readout electronics uses a
flash ADC (FADC) system with 100 ns period to record the phototube
information.  

We analyze our data in two ways: in monocular mode we analyze each of
our detectors' data independently.  This mode gives us the best
statistics and the widest energy range: from $10^{17.2}$ to
$10^{20.5}$ eV.  In analyzing the two detectors' data together, called
stereoscopic mode, we get the best resolution, but have fewer events,
and cover a narrower energy range: $10^{18.5}$ to $10^{20.5}$ eV.  In
this paper we consider our data analyzed in monocular mode.  Our
stereo spectrum agrees very well with the spectrum being presented
here, both in shape and in normalization.

\section{Calibration and Corrections}

The calibration of our detectors in terms of photons is performed by
carrying a very stable xenon flash lamp to each mirror on a monthly
basis, and illuminating the phototube cluster.  Two methods of
analysis of the xenon lamp data, using the absolute intensity and
using photon statistics, agree within uncertainties.  Ultimately our
calibration comes from NIST-traceable photodiodes.  We check the
calibration with hybrid photodiodes and laser shots fired from the
field in the aperture of our detectors.  Night-to-night relative
variations are monitored with a YAG laser.  We achieve 10\% accuracy
in our optical calibrations.

The atmosphere is not only our calorimeter, it also is the medium
through which fluorescence light is transmitted to our detectors.  We
monitor the transmission of the atmosphere by two methods.  Raleigh
scattering from the molecular portion of the atmosphere varies little
with time.  We monitor this through the density of the atmosphere,
which is measured by radiosonde balloons from two neighboring
airports.  To measure scattering from the aerosol component of the
atmosphere we fire lasers into the sky and observe the scattered light
with our detectors.  A series of laser shots that takes 50 minutes to
perform is begun every hour.  We measure the vertical aerosol optical
depth (VAOD) and the horizontal extinction length and phase function.
The average VAOD is about 0.04, with an rms of 0.02.  The skies at our
site are very clear.  An event 25 km distant has an aerosol correction
to its energy of about 10\% on average.  Since 2.5 years of our early
data was collected before our lasers were constructed, and we want to
have a consistent analysis, the spectra presented here are calculated
using a constant-atmosphere assumption, using the average value of
VAOD.  We have tested this assumption by calculating the spectrum from
our later data, using the actual hourly measurements.  We get the same
answer to within a few percent
\cite{andreas}.

The yield of fluorescence photons per minimum ionizing particle per
meter of path length is a parameter that enters directly into our
calculation of cosmic rays' energies.  It has been measured previously
\cite{kaknag461}.  The three measurements in reference
\cite{kaknag461}  agree very well, and a fit to the three measurements
indicates that the absolute flux is known to $\pm 6\%$.

\section{Data Analysis}

The analysis begins with a pattern recognition step.  Tubes on a track
are identified through being contiguous with other tubes in position
and time.  The plane that passes through the shower and the detector
is found.  We make a plot of the time each tube fires vs the angle of
the tube in the shower-detector plane.  This distribution is not a
straight line: the curvature arises due to the geometry of the
detector and shower.  This curvature is then used in a fit to
determine the event geometry.  We measure the perpendicular distance
to the shower and the angle of the shower in the shower-detector
plane.  This angle is measured to about $5^{\circ}$ accuracy for the
HiRes-II detector.

\begin{figure}[ht]
\begin{center}
\mbox{  \epsfysize 3.in \epsfbox[ 0 0 567 567 ]
        { 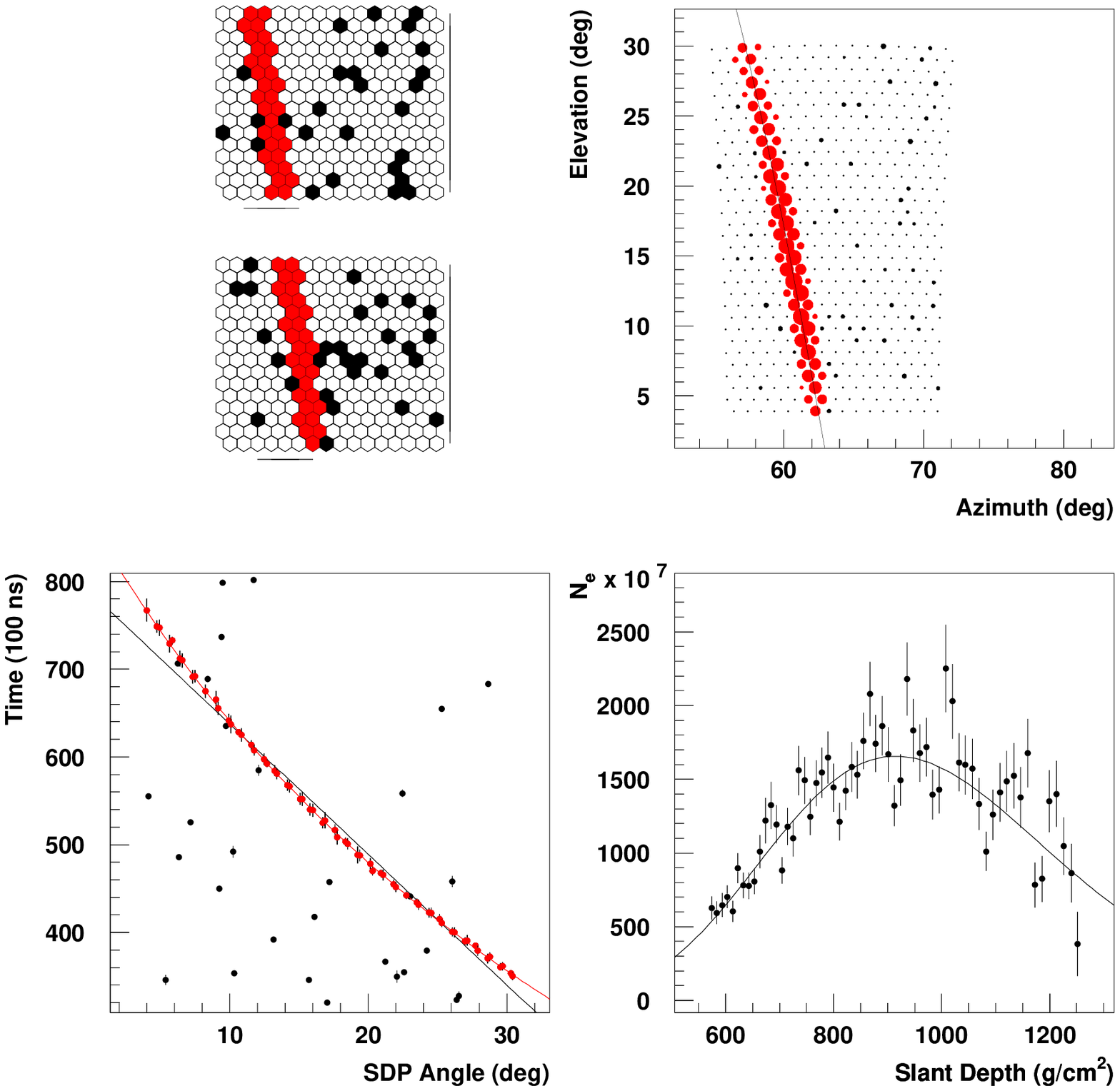 } } 
\mbox{  \epsfysize 2.7in \epsfbox[ 80 187 493 576 ]
        { 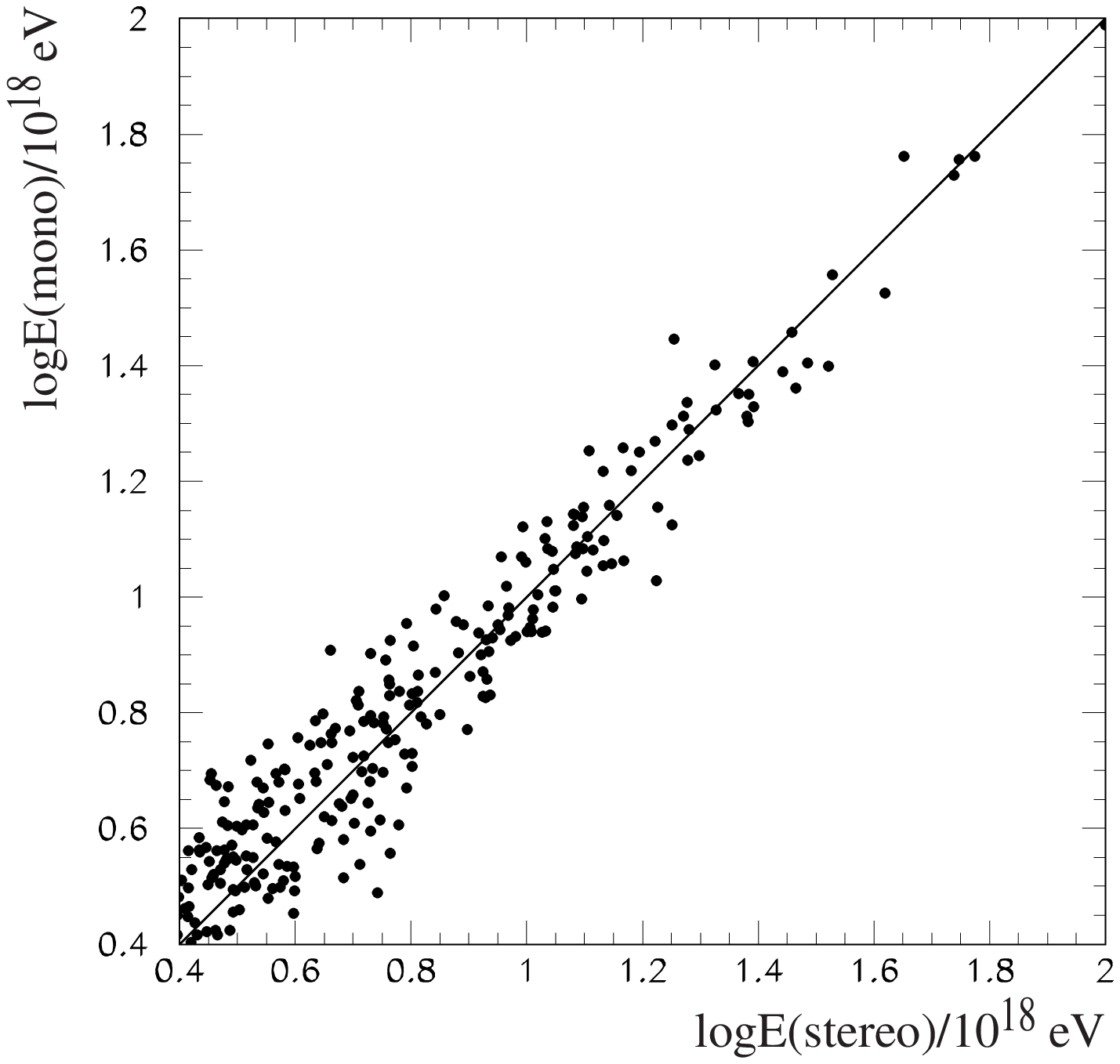 } } 
\end{center}
    \caption[]{ Left figure: Illustration of monocular event
reconstruction for one event seen by the HiRes-II detector.  The two
mirrors that saw the cosmic ray track, all the tubes' elevation vs
azimuthal angles, the time of tubes firing vs their angles along the
shower-detector plane, and the profile with its Gaisser-Hillas fit are
all shown.  
Right figure: HiRes-I energies calculated with the event
geometry reconstructed in stereo, vs. the energy calculated when the
geometry is determined using profile-constrained fit. }
\label{fig:mono}
\end{figure}

The next step is to calculate the number of charged particles in the
shower as a function of the slant depth.  This is called the profile
of the shower.  A fit to the profile is performed using the
Gaisser-Hillas function \cite{g-h}, from which the depth of shower
maximum is determined.  The integral of the Gaisser-Hillas function is
used to determine the calorimetric energy of the shower.  The left
part of Figure \ref{fig:mono} illustrates monocular event
reconstruction for the HiRes-II detector.

Because tracks seen in the HiRes-I detector are shorter, in the fit
for the geometry we also use the shape of the reconstructed profile as
a constraint.  Again the Gaisser-Hillas function shape is used.  An
accuracy of about $7^{\circ}$ in the in-plane angle is achieved in
this fit.  The right half of Figure \ref{fig:mono} shows the energy of
events seen by the HiRes-I detector reconstructed using the profile
constrained fit vs the energy when the event geometry is reconstructed
stereoscopically (with about $1^{\circ}$ uncertainty in the in-plane
angle).  It is clear that the profile-constrained fit works very well
in reconstructing events' energies.

Finally a correction is made for missing energy: for neutrinos and
muons that strike the earth and whose energy is not counted in the
calorimetric energy.  This correction is determined from shower
simulations using Corsika \cite{corsika} and the hadronic generator
program QGSJet \cite{qgsjet}.  This correction is typically about
$10\%$.  The hadronic generator program Sibyll \cite{sibyll} predicts
the same missing energy correction to 2\%.

\section{Aperture Calculation}

To calculate the aperture of the experiment we perform a full Monte
Carlo (MC) simulation, including shower development (using actual
Corsika showers), fluorescence and Cerenkov photon generation,
propagation through the atmosphere, light collection by our detectors,
and the response of our trigger and readout electronics.  The MC
events are written out in exactly the same format as the data, and are
analyzed using the same programs.  Previous measurements of the
spectrum and composition by the Fly's Eye Experiment \cite{flyseye},
the HiRes/MIA hybrid experiment \cite{hiresmia}, and HiRes stereo (for
composition) \cite{greg} are used in the simulation.

\begin{figure}[ht]
\begin{center}
\mbox{  \epsfysize 3.in \epsfbox[ 59  205  559  735 ]
        { 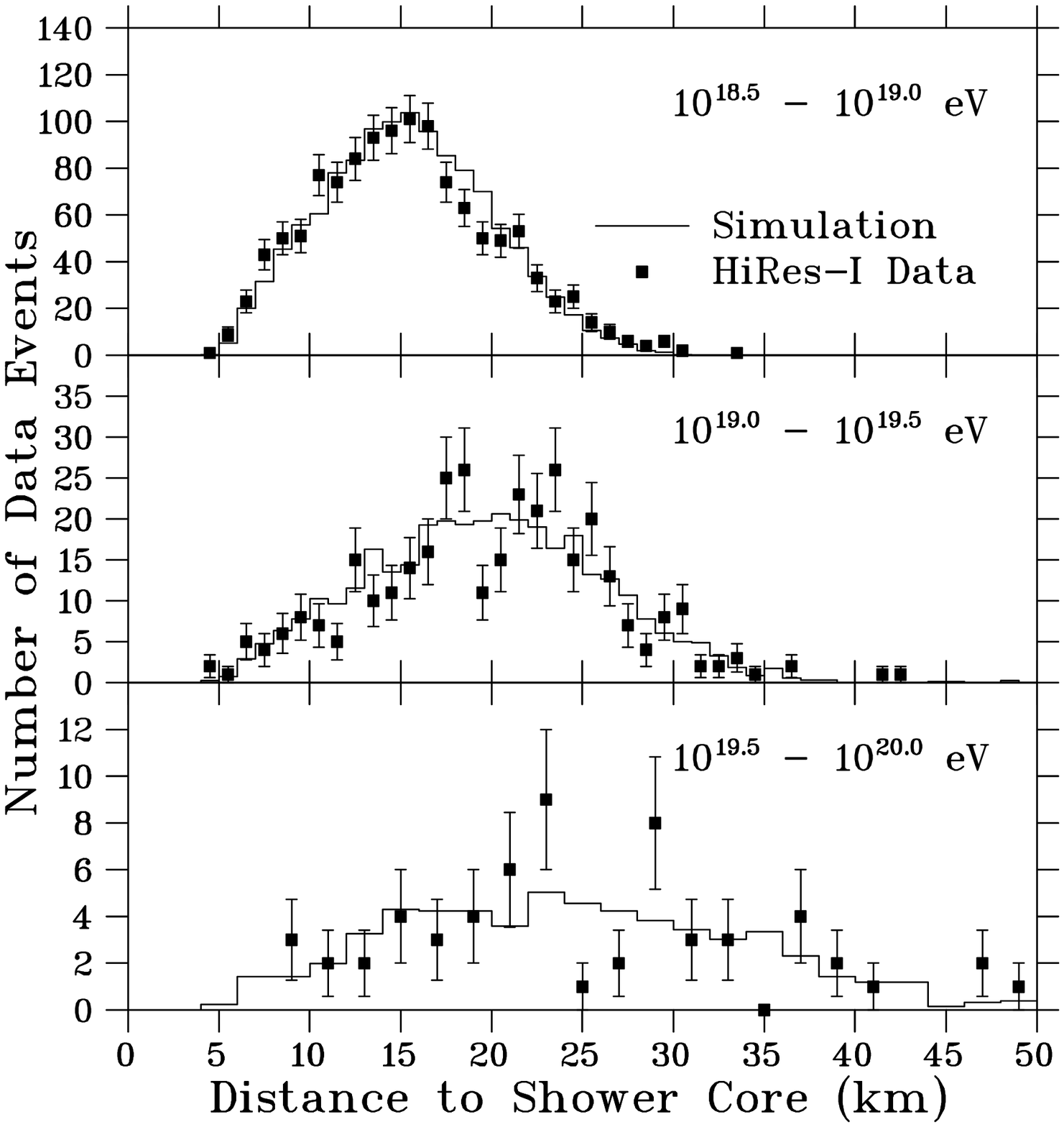 } } 
\mbox{  \epsfysize 3.in \epsfbox[ 0 130 567 697 ]
        { 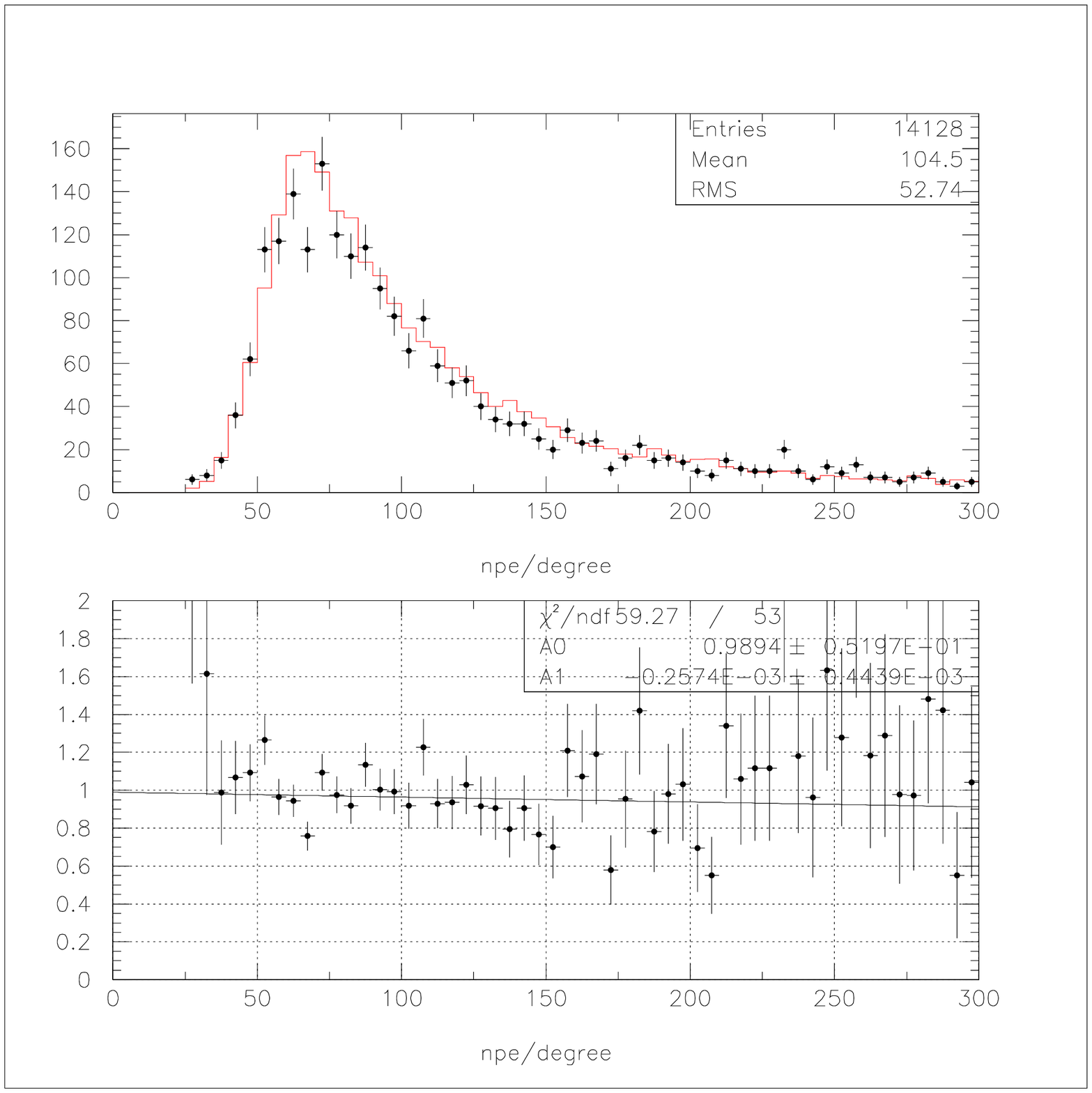 } } 
\end{center}
    \caption[]{ Left figure: Histograms comparing the distance to the
shower core in Monte Carlo and data for the HiRes-I detector.  Three
energy bins are shown.
Right figure: Histogram showing the comparison of Monte Carlo to data
for the number of photoelectrons per degree of track for events seen
by the HiRes-II detector.  The lower part of the figure is the
bin-by-bin ratio of data events to the Monte Carlo events.}
\label{fig:comps}
\end{figure}

The way of judging the success of such a simulation is by comparing MC
histograms of kinematic and physics variables with those of the data.
If they are identical then one says that one understands one's
experiment.  Our MC simulates the data very well.  Two comparison
plots with excellent agreement between MC and data, typical of many,
are shown in Figure \ref{fig:comps}.  The left part of the figure
shows the distance to the shower core for the HiRes-I detector, and
the right part shows the number of photoelectrons per degree of track
for events seen by the HiRes-II detector, showing that the same amount
of light comes from the sky in the MC as in the data.

\begin{figure}[tbh]
\epsfysize=3.5in
\centerline{\epsffile{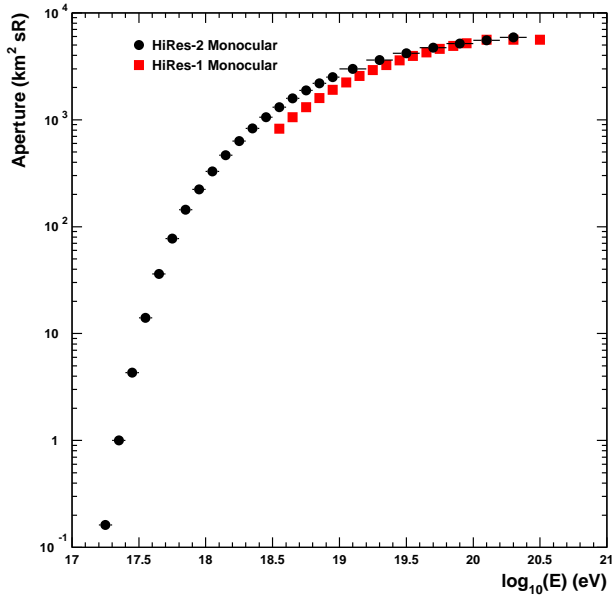}}
\caption{The aperture of the HiRes detectors operating in monocular
mode.  Apertures of HiRes-I and HiREs-II are shown.}
\label{fig:aperture}
\end{figure}

Figure \ref{fig:aperture} shows the result of the aperture
calculation.  The apertures of the HiRes-I and HiRes-II detectors are
shown.

\section{The Spectrum}

\begin{figure}[tbh]
\epsfysize=4.0in
\centerline{\epsffile{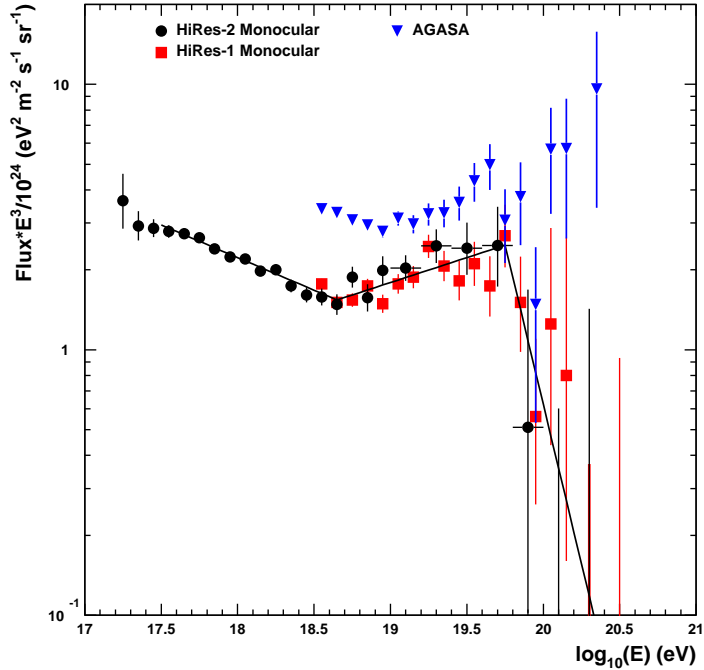}}
\caption{The cosmic ray spectrum measured of the HiRes detectors
operating in monocular mode.  The spectrum of the HiRes-I and HiRes-II
detectors are shown.  The spectrum of the AGASA experiment is also
shown.  The falling-off of the HiRes spectrum above
$10^{19.8}$ eV is the GZK cutoff, and the dip at $10^{18.6}$ eV is the
``ankle''.}
\label{fig:spectrum}
\end{figure}

Evidence for the suppression of the flux of cosmic rays at high
energies has been published in previous spectrum measurements by our
collaboration \cite{pastspectra}.  Figure \ref{fig:spectrum} shows the
current monocular spectra of the two HiRes detectors.  The flux times
$E^3$ is shown.  The HiRes-I data were collected from May, 1997 to
May, 2005, and that from HiRes-II from December, 1999 to August,
2005.  

The GZK cutoff stands out clearly as the suppression of the spectrum
at $10^{19.8}$ eV.  The dip at $10^{18.6}$ eV is a feature known as
the ``ankle''.  Fits to the spectrum \cite{fits} show that the ankle
is likely caused by $e^+e^-$ pair production in the same interactions
between CMBR photons and cosmic ray protons where pion production
produces the GZK cutoff.  Previous experiments have observed a feature
known as the ``second knee'' \cite{pravdin,hiresmia,secondknee} near
$10^{17.6}$ eV, although there is considerable uncertainty in this
energy.  The statistical power of our data is poor in this energy
range, and below $10^{18}$ eV systematic uncertainties are growing as
well.  So HiRes cannot claim to observe the second knee.

At lower energies, the spectrum of cosmic rays can be fit to power
laws over ranges of energy.  This is also true for the spectrum in
Figure \ref{fig:spectrum}.  Superimposed on the spectrum are three
lines fit to the data with break points between the line segments also
found by the fitting program.  There are two breaks found, at
$\log{E(eV)}$ of $19.75 \pm 0.05$ for the GZK cutoff, and $18.65 \pm
0.05$ for the ankle.  The $\chi^2$ of this fit is 34.7 for 35 degrees
of freedom.  A fit with only one break point allowed finds the ankle,
and the $\chi^2$ is 68.2 for 37 degrees of freedom.  Correcting for
the number of degrees of freedom, the difference in $\chi^2$ of 31.5
corresponds to about a 5 1/2 standard deviation statistical
significance to our observation of the GZK cutoff.

Another way of calculating the statistical significance of the break
at $10^{19.8}$ eV is to calculate the number of events that we should
have collected if there were no break, and compare it with the true
number of events.  Using our exposure we should have seen 51.1 events
if there were no break, but we actually observed 15.  The Poisson
probability of seeing 15 or fewer with a mean of 51.1 is $3 \times
10^{-9}$, which corresponds to 5.8$\sigma$, consistent with the
$\chi^2$ calculation in the previous paragraph.  Removing the
overlapping events seen by both HiRes-I and HiRes-II and redoing the
calculation yields a 5.2 standard deviation effect.

\section{A ``Test Beam'' of High Energy Events}

\begin{figure}[tbh]
\epsfysize=4.0in
\centerline{\epsffile{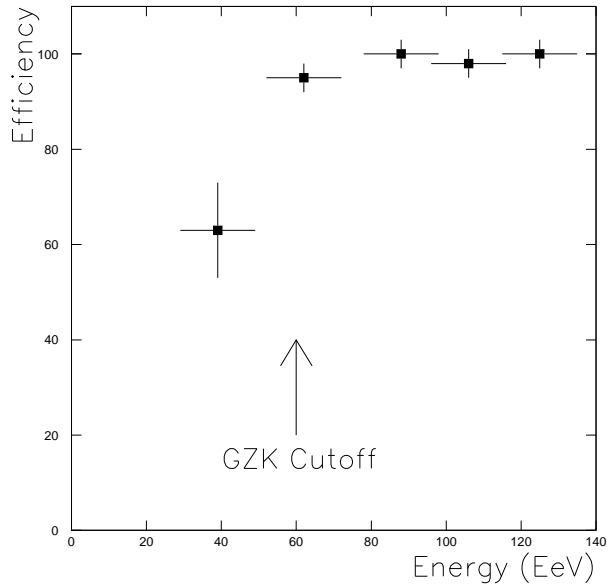}}
\caption{The efficiency for triggering on and reconstructing Terra
Ranch laser shots by the HiRes-II detector.  The five laser energies
are plotted on the ordinate at the equivalent energy a cosmic ray
shower of that brightness would have.  This figure proves that HiRes
has excellent sensitivity to cosmic ray showers at and above the GZK
energy, even though those showers are not present in our data; i.e.,
the absence of these events is due to the GZK cutoff, and is not an
instrumental effect.}
\label{fig:terra}
\end{figure}

Since we do not observe a continuation of the flux of cosmic rays
above the GZK energy one might ask whether our experiment is capable
of seeing these events.  To answer this question we constructed a
``test beam'' of high energy tracks using a laser.  This laser is
located at Terra Ranch, 35 km from the HiRes-II detector, at the edge
of our aperture, and it fires a beam of light vertically at a
wavelength of 355 nm.  The atmosphere scatters the light into the
aperture of our detectors, and we trigger on it just like a cosmic ray
shower.  We have measured the efficiency for triggering on the
scattered light and reconstructing it as we would do for a shower, and
this efficiency is shown in Figure \ref{fig:terra}.  The data for this
figure was collected on good-weather nights where the cosmic ray data
were used in our spectrum measurement.  The laser fires at five energy
levels.  The specific brightness of a cosmic ray shower, for example
measured in photoelectrons per degree of track length, is
characteristic of its energy, and we have plotted the five laser
energies in Figure \ref{fig:terra} at the equivalent cosmic ray
energies.  Figure \ref{fig:terra} proves that we have excellent
sensitivity to cosmic ray showers at and above the energy of the GZK
cutoff.  The fact that super-GZK events are absent in our data is thus
due to the GZK cutoff itself, and is not an instrumental effect.

\section{Systematic Uncertainties}

The main contributions to the systematic uncertainty in the HiRes
energy scale and flux measurements are: phototube calibration (10\%),
fluorescent yield (6\%), missing energy correction (5\%), aerosol part
of atmospheric correction (5\%), mean dE/dx\footnote{The two newest
versions of QGSJet yield mean dE/dx estimates for showers that differ
by 10\%.} (10\%), for a total energy scale uncertainty of 17\%, and an
uncertainty in the flux of 30\%.

\section{Summary}

We have measured the flux of ultrahigh energy cosmic rays by the
fluorescence technique, over the energy range $10^{17.2}$ to
$10^{20.5}$ eV.  We observe the GZK cutoff and the ankle of the
cosmic ray spectrum.  We cannot claim to observe the second knee.  The
statistical power of our observation of the GZK cutoff is just over
$5\sigma$.  The systematic uncertainty in our energy scale is about
17\%, and the uncertainty in the flux is about 30\%.

\section{Acknowledgements}

This work is supported by US NSF grants PHY-9321949, PHY-9322298,
PHY-9904048, PHY-9974537, Phy-0098826, PHY-0140688, PHY-0245428,
PHY-0305516, PHY-030098, and by the DOE grant FG03-92ER40732.  We
gratefully acknowledge the contrinbutions from the technical staffs of
our home institutions.  The cooperation of Colonels E. Fischer and
G. Harter, the US Army, and the Dugway Proving Ground staff is greatly
appreciated.

\end{document}